# Extended multi-phase gas reservoirs in the z=4.3 protocluster SPT2349-56: non-stellar ionisation sources?

Kevin C. Harrington[1, 2, 3, 4], Amit Vishwas[5], Allison W. S. Man[6], Carlos De Breuck[7], Padelis P. Papadopoulos[8, 9, 7, 10], Paola Andreani[7], and Thomas. G. Bisbas[11]

[1] Joint ALMA Observatory, Alonso de Córdova 3107, Vitacura, Casilla 19001, Santiago de Chile, Chile
    e-mail: kevin.harrington@alma.cl
[2] National Astronomical Observatory of Japan, Los Abedules 3085 Oficina 701, Vitacura 763 0414, Santiago, Chile
[3] European Southern Observatory, Alonso de Córdova 3107, Vitacura, Casilla 19001, Santiago de Chile, Chile
[4] Instituto de Estudios Astrofísicos, Facultad de Ingeniería y 455 Ciencias, Universidad Diego Portales, Av. Ejército Libertador 441, Santiago, Chile
[5] Cornell Center for Astrophysics and Planetary Sciences, Cornell University, Ithaca, NY, 14853, USA.
[6] Department of Physics & Astronomy, University of British Columbia, 6224 Agricultural Road, Vancouver BC, V6T 1Z1, Canada
[7] European Southern Observatory, Karl-Schwarzschild-Strasse 2, 85748 Garching, Germany
[8] Department of Physics, Section of Astrophysics, Astronomy and Mechanics, Aristotle University of Thessaloniki, 54124 Thessaloniki, Greece
[9] Research Center for Astronomy, Academy of Athens, Soranou Efesiou 4, 11527 Athens, Greece
[10] Max Planck Institute für Astrophysik, Karl-Schwarzschild-Strasse 1, 85748 Garching, Germany
[11] Research Center for Astronomical Computing, Zhejiang Laboratory, Hangzhou, 311000, China

**ABSTRACT**

We aim to characterize the multi-phase gas in the SPT2349-56 protocluster at z=4.3, known to host one of the most starbursting and AGN-rich high redshift environments. For this purpose we conducted APEX single dish observations of the [CII] 158 $\mu$m (hereafter [CII]) line towards the Core and North components, previously imaged with the ALMA 12-m array. We also present the first [OIII] 88 $\mu$m (hereafter [OIII]) line observations in such high redshift protocluster system. We obtain a [CII] line luminosity ~1.7× more than the one recovered by ALMA towards the Core, while remarkably we recover 4× more [CII] line emission than the one found in deep ALMA images towards the North component, suggesting that the most massive gas reservoirs lie in the less extreme regions of this protocluster system. A minimum ionised gas mass of $M_{min}(H^+) \sim 3.7 \times 10^{10}$ $M_\odot$ is deduced from the [OIII] line, amounting to 30% of the molecular gas mass in the same area, indicating that a full map of the cluster is necessary for determining the large-scale value. Finally we obtain star formation rate (SFR) estimates using the [OIII] line luminosity, and the corresponding ionised gas mass. These yield values that can surpass the far-IR continuum-derived SFR (under the assumption of a standard stellar IMF), which can be reconciled only if non-stellar ionising sources contribute to the [OIII] line luminosity, or a top-heavy stellar IMF produces a larger fraction of O stars per total stellar mass, a distinct possibility in High-Energy-Particle (HEP) rather than (UV-photon)-dominated environments in clusters. Future work using far-IR fine-structure and molecular/neutral-atomic lines is necessary for determining the thermal/ionisation states of the multi-phase medium in this protocluster, what maintains them, and for resolving the apparent SFR discrepancy. These line ratios must be measured over a wide range of spatial scales, from individual galaxies up to Circumgalactic and Intercluster Medium (CGM/ICM) scales, which ultimately requires combining wide-field single-dish and high resolution interferometric observations of such lines in protocluster environments where HEP- and UV-dominated ISM phases can co-exist.

## 1. Introduction

Distant "protoclusters", the precursors to modern-day galaxy clusters, are unique laboratories of cosmological structure formation and galaxy evolution across cosmic time (Overzier 2016; Alberts & Noble 2022). Galaxy assembly and star formation both in-situ (within galaxies) and ex-situ (within dark matter haloes but beyond the luminous extent of galaxies) in protoclusters require mechanisms to effectively cool the gas in order to aid its gravitational collapse. Importantly, unlike the interstellar medium (ISM) within galaxies, the thermal/ionisation states of the intracluster/circumgalactic medium (ICM/CGM) gas phases in clusters may not be far-UV radiation-controlled, but High Energy Particle (HEP)-dominated (Ferland et al. 2008, 2009; Lim et al. 2017). Unfortunately it is also in such HEP-dominated environments where CO, the main tracer of $H_2$, can be effectively destroyed, yielding [CI] and [CII]-rich gas phases (Bisbas et al. 2015, 2017, 2021, 2025), with CO remaining to ″mark″ only denser sub-regions surrounded by otherwise (CO-poor)/([CI]/[CII])-rich $H_2$ gas reservoirs (Papadopoulos et al. 2018). A C-rich/CO-poor circumgalactic medium (CGM) $H_2$ gas reservoir as extended as ≈50 kpc has already been found in the protocluster around the Spiderweb galaxy at z=2.2 (Emonts et al. 2016, 2018) and was recently confirmed to be embedded in a hot intracluster medium (ICM) via the Sunyaev Zel'dovich Effect (Di Mascolo et al. 2023). Furthermore, cold $H_2$ gas streams traced by the [CI] (1-0) line emission, extending across 100 kpc, have been shown fueling 4C 41.17, a massive, radio galaxy at z=3.8 (Emonts et al. 2023). Thus there is a clear need to study the colder (T~(20-100) K) gas phase expected alongside the warmer (~$10^4$-$10^6$ K) ICM/CGM gas of protoclusters, with the former being the readily available fuel of their starburst activity in the Early Universe.

Far-IR fine-structure lines can dominate the cooling energy budget within the cold neutral/ionised ISM (see e.g., Spinoglio et al. 2012; Fernández-Ontiveros et al. 2016), with [CII] often considered the dominant coolant, tracing both neutral and ionised gas (Stacey et al. 2010; Vallini et al. 2015; Gullberg et al. 2015; Lagache et al. 2018; Cormier et al. 2019). Nevertheless, [OIII] 88 $\mu$m ([OIII] thereafter) line emission is often just as bright, while its high ionisation potential (IP) sets it up for tracing solely the fully ionised gas phase[1]. In the ISM of galaxies the [OIII] line emission comes from compact (and thus young) HII regions around massive stars (Cormier et al. 2012; Vallini et al. 2017; Vishwas et al. 2018; Arata et al. 2020), yet it has been difficult to observe even at high redshift due to poor atmospheric transmission at its wavelength. Space-based observations of [CII] and [OIII] in local metal-poor systems yielded $L_{[OIII]}/L_{[CII]}$~1 (Madden et al. 2013), even as such line ratios are extremely ambiguous in their possible interpretations as the two lines trace vastly different thermal and ionisation states. Indeed [CII] dominates the cooling budget of the neutral ISM in many local starbursts (De Looze et al. 2014; Díaz-Santos et al. 2017) (even as HII regions and a diffuse low-ionisation phase can make ~10% contribution to its luminosity in metal-poor ISM), yet given the low IP value of neutral carbon (~11.3 eV), it could not trace the same phase as the [OIII] line (~35.1 eV for $O^{++}$). Indeed with a higher IP where [OIII] is abundant, Carbon would not remain in a singly ionised state (the IP for $C^+$ is ~24.4 eV). Standard photoionisation codes such as CLOUDY (Ferland et al. 2017) naturally yield a multiplicity of thermal/ionisation states within their 1-D gas columns, irradiated by stellar or AGN spectral energy distributions, yet they certainly need more than a single line ratio as an input to give any useful results.

Both the [CII] and [OIII] lines become accessible to observations in one of the most distant protocluster systems found, namely the SPT2349-56 at $z$ = 4.304. It was originally detected in the South Pole Telescope (SPT) 2500 deg$^2$ millimeter(mm)-survey (Vieira et al. 2010; Mocanu et al. 2013), and spectroscopically confirmed by Strandet et al. (2016) using CO(4-3) and [CII] from the APEX telescope after follow-up of the brightest 870$\mu$m detections with the Large APEX BOlometer CAmera (LABOCA; Kreysa et al. 2003; Siringo et al. 2009). LABOCA dust continuum maps initially revealed a central source and a northern extension, and subsequent ALMA observations found ~29 CO(4-3)/[CII] -emitting objects spread over a few hundred kpc (proper distance) towards the core and northern extension, with an estimated global star formation rate of SFR~$10^4$ $M_\odot$yr$^{-1}$(Miller et al. 2018; Hill et al. 2020, 2022). Here we present new [CII] and [OIII] observations in SPT2349-56, sensitive to large-scale gas distributions in and around the protocluster members, where previous interferometric measurements with ALMA of [CII] in the Core and North may have resolved out the corresponding line emission, as suggested by Zhou et al. (2025). We present the APEX and ALMA measurements in Sect. 2, our results and discussion are presented in Sect. 3, followed by a brief outlook in Sect. 5 that further motivates the need for sensitive single-dish measurements to study such protocluster systems at high redshift. Throughout this work, we adopt a standard $\Lambda$CDM cosmology: $H_0$ = 70 km s$^{-1}$ Mpc$^{-1}$, $\Omega_m$ = 0.3 and $\Omega_\Lambda$ = 0.7.

## 2. Observations and data reduction

### 2.1. APEX [CII] and [OIII] line measurements

The Atacama Pathfinder EXperiment (APEX) telescope is a 12-m single-dish telescope in the Chajnantor plateau in the Atacama desert of northern Chile. APEX observations targeted both the [CII] and [OIII] emission lines using the known redshift $z$ = 4.304 and coordinates as first discovered in Strandet et al. (2016) and Miller et al. (2018). The "Core" pointing in the observations corresponds most closely to object "C1" in Hill et al. (2020, Core: RA:23:49:42.65, DEC:-56:38:19.4) . We also report novel observations of [CII] line emission from the "North" system located ~45" away (North; RA: 23:49:42.53, DEC: -56:37:33.2). Figure 1 shows the APEX full-width-at-half-maximum (FWHM) of the primary beam of each pointing overlaid on the dust continuum emission from ALMA band 6 / 1mm measurements (Sect. 2.2) for both the Core and North. The [OIII] 88 $\mu$m measurements are the first observations of this fine-structure line in the Core. Both lines are observable at favourable atmospheric windows at ~358 GHz and ~640 GHz for each line. The [CII] and [OIII] lines were observed with the SEPIA345 (Meledin et al. 2022) and SEPIA660 (Baryshev et al. 2015) instruments [2] respectively, between 2024 August 31 and September 6 (Project ID: C-0114.F-9704-2024; PI: K. Harrington) and at night during stable conditions. Precipitable water vapour (PWV) varied between ~0.5-0.8 mm for SEPIA345 (Meledin et al. 2022), the single-pixel, dual-polarization, 2-sideband (SB; 4-GHz per upper and lower sideband) heterodyne receiver. Excellent observing conditions with PWV <0.5mm enabled higher frequency observations with SEPIA660 (Baryshev et al. 2015), which has the same receiver design as SEPIA345. All observations were performed within the APEX control system (Muders et al. 2006), data were recorded using the MPIfR eXtended bandwidth Fast Fourier Transform spectrometers (FFTS; Klein et al. 2006), and each of the scans reduced and analysed using GILDAS: Grenoble Image and Line Data Analysis Software[3]. The baseline stability depends strongly on the observed frequency and/or weather conditions. We first subtract a first-order baseline from the emission line-free channels, guided by the known ~ 1600 km s$^{-1}$ extent of the line profile (Miller et al. 2018; Hill et al. 2020), in the native resolution. We smoothed the baseline-subtracted spectrum before averaging to produce the final spectrum, obtaining a root-mean-square (RMS) sensitivity of 0.3–0.6 mK (at 90–110 km s$^{-1}$ channel resolution) at ~358 GHz, and 0.3–0.4 mK (at 110–150 km s$^{-1}$ channel resolution) at ~640 GHz. Using the APEX telescope efficiencies, 46.4 Jy K$^{-1}$ and 59.5 Jy K$^{-1}$ at 358.4 GHz and 639.7 GHz, respectively[4], we scale to a measured flux density. We then integrate across each line profile to calculate line luminosities (Table 1; see e.g. Solomon & Vanden Bout 2005; Carilli & Walter 2013 and references therein for formalism used). The total uncertainties of around 15–20% come primarily from the flux calibration, pointing/focus, and baseline subtraction errors.

---

[1] Excitation potential $E_{ul}/k_B$ ~91 K and ~163 K, and critical densities of $n_{crit}(H) = 3 \times 10^3$ cm$^{-3}$ (for HI and $H_2$ as collisional partners) and n(e) = 510 cm$^{-3}$ for the [CII] and [OIII] lines, respectively.

[2] APEX Instruments Overview: https://www.apex-telescope.org/ns/observing-run/observing/the-telescope/instruments/

[3] Software information can be found at: http://www.iram.fr/IRAMFR/GILDAS.

[4] https://www.apex-telescope.org/telescope/efficiency/index.php

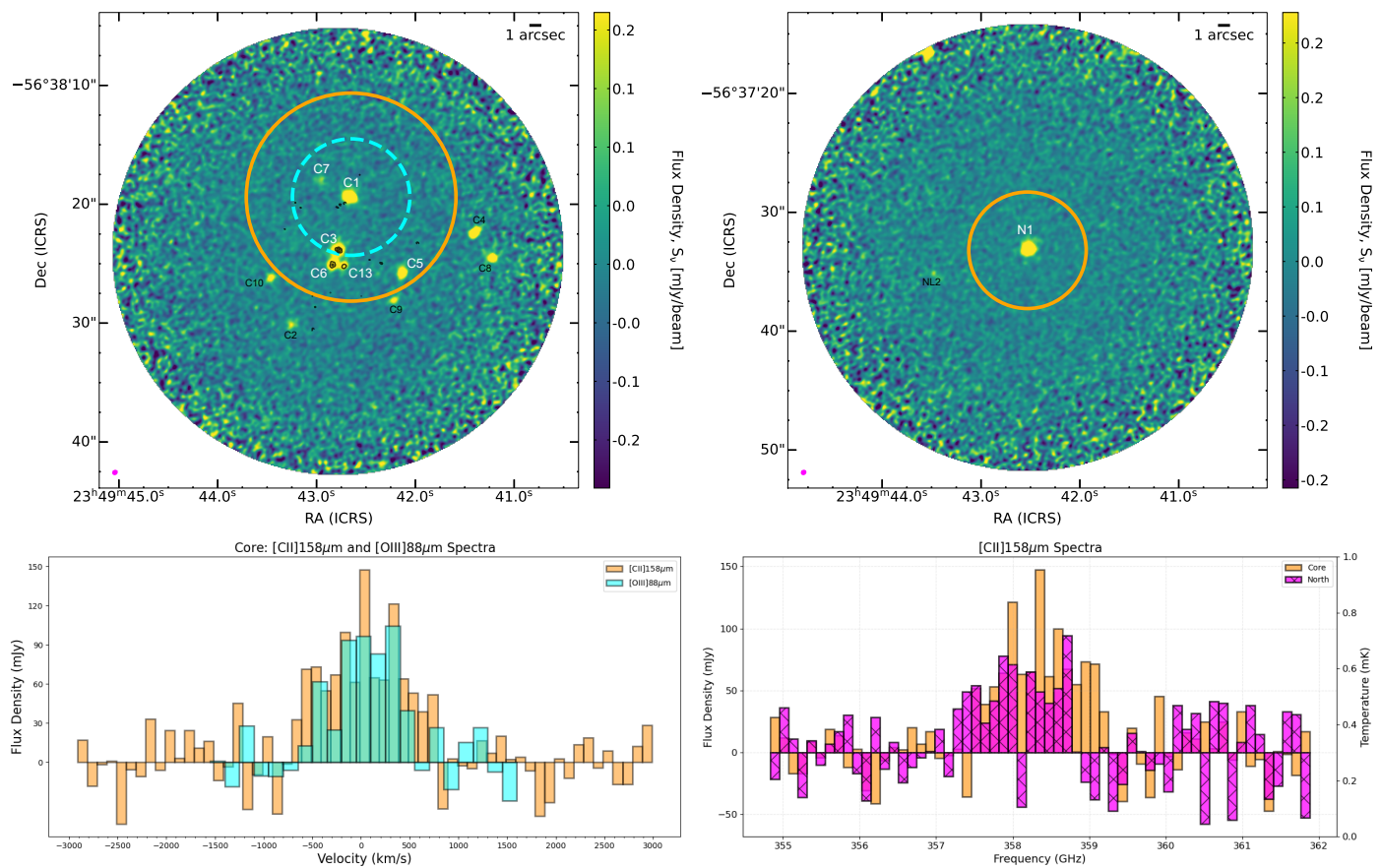

**Fig. 1.** ALMA band 6 dust continuum emission from SPT2349-56, Core region (Top Left) and the North extension (Top Right) 45″ away. Pointing center and instrument FWHM of APEX SEPIA345 (orange) and SEPIA660 (cyan) observations are overlaid. Band 6 synthesized beam (magenta) is shown in the bottom left of each panel. The black contours show ALMA band 10 dust continuum detections towards the Core region at signal-to-noise ratio levels of 3,6,10,15. Several continuum sources are labeled as identified by Hill et al. (2020). Bottom Left: The [CII] (orange; $\delta V = 110$ km s$^{-1}$) and [OIII] 88 $\mu$m (cyan; $\delta V = 140$ km s$^{-1}$) spectra (Flux density vs velocity, corrected using $z = 4.304$). Bottom Right: The [CII] spectra (Flux density / antenna temperature vs frequency) shown for both observations of the Core (orange; $\delta V = 100$ km s$^{-1}$) and North (fuchsia; $\delta V = 100$ km s$^{-1}$) fields.

**Table 1.** APEX single-dish measurements of [CII] and [OIII]

| Line | Integrated Flux (Jy km s$^{-1}$) | $L_{\rm line}$ ($10^{10}$ L$_\odot$) | $L'_{\rm line}$ ($10^{10}$ K km s$^{-1}$ pc$^2$) | RMS per 100 km s$^{-1}$ (mK) |
|---|---|---|---|---|
| [CII] Core | 101.7 ±15.1 | 6.1 ±0.9 | 27.5 ±4.1 | 0.38 |
| [OIII] Core | 78.2 ±15.6 | 8.3 ±1.7 | 6.6 ±1.3 | 0.65 |
| [CII] North | 80.4 ±12.1 | 4.8 ±0.7 | 21.8 ±3.3 | 0.3 |

### 2.2. ALMA band 6 and band 10 dust continuum

Here we utilise ALMA dust continuum measurements to identify the protocluster galaxy members with respect to the [CII] and [OIII] line emission measurements from the APEX detections. Band 10 (Uzawa et al. 2013) dust continuum measurements were recently obtained at 875–900 GHz (PID 2024.1.01465.S; PI: K. Harrington) from ALMA 12-m observations between 2024 October 12 and November 6 under excellent conditions: precipitable water vapor and phase RMS 0.3 mm and $< 15\mu$m, respectively. We obtained calibrated measurement sets of the two observations by re-running the ALMA pipeline on the raw data, using CASA (McMullin et al. 2007). We then inspected the pipeline weblog and concluded that sufficient flagging had been applied, before imaging using the tclean task in CASA with all spectral windows and an auto-multithresh (Kepley et al. 2020) cutoff, resulting in an RMS $\sim 0.7$ mJy beam$^{-1}$. Three objects have been detected, with more observational details described in Appendix Sect. A (Table A.1). We also used archival ALMA 12-m imaging from observations conducted on 2022 September 1 of the Core and North in band 6 (PID 2021.1.01313.S; PI: R. Canning). We utilise the pipeline-products available from the ALMA archive to plot a continuum image from 223–240 GHz (Fig. 1).

## 3. Results

### 3.1. Excess [CII] line emission

The SPT2349-56 protocluster at z~4.3 consists of multiple member galaxies that are concentrated in a Northern and Southern spatial configuration (Fig. 1, Fig B.1). The APEX [CII] observa-tion towards the Core suggests these single-dish measurements capture the [Cɪɪ] line emission from 19 out of 23 objects presented in deep ALMA 12-m observations (∼0.2 mJy beam$^{-1}$ per 13 km s$^{-1}$ channel resolution) presented in Hill et al. (2020). The sum of the [Cɪɪ] line fluxes for all 19 objects as detected with ALMA 12-m array, excluding objects 'C2,C4,C8,C19'[5], results in $\int_{\Delta V} S_\nu([CII])dV$∼60 Jy km s$^{-1}$(Hill et al. 2020). To evaluate whether ALMA interferometric observations have filtered out extended [Cɪɪ] beyond the maximum resolvable scale (∼5-6") and sensitivities probed by both Miller et al. (2018) and Hill et al. (2020), we compare the individual [Cɪɪ] line fluxes obtained from interferometry and the single-dish observations of the Core. The APEX [Cɪɪ] measurements (Table 1) indicate more than 100 Jy km s$^{-1}$, i.e. about 1.7× the line emission compared to ALMA. Based on the discussion of the luminosity function in Hill et al. (2020) and the sensitivity of ALMA observations being compared with, we estimate that no more than 10-15% of the excess flux can be attributed to low-luminosity galaxies that are within the (relatively small) beam, but undetected with ALMA.

In addition, we present the first single-dish observation towards the North (Table 1), centered on object 'N1' presented in Hill et al. (2020). Our North pointing consists of a single central object in the archival band 6 continuum and [Cɪɪ], and Hill et al. (2020) report that this one object ('N1') has $\int_{\Delta V} S_\nu([CII])dV$=18.9 Jy km s$^{-1}$. We measure a [Cɪɪ] line flux of 80 Jy km s$^{-1}$ from APEX observations, i.e., 4× the amount of [Cɪɪ] line emission than measured by the deep ALMA observations. This indicates that, indeedamente, the largest non-galaxy-bound reservoir of [Cɪɪ]-emitting gas in the SPT2349-56 protocluster lies in the North. Moreover, the North component [Cɪɪ] emission amounts to nearly 80% of that of the Core, yet it contains only one [Cɪɪ]-line emitter. The combined Core and North [Cɪɪ] line luminosity (∼ 11 × 10$^{10}$L$_\odot$ amounts ∼0.14% of the total IR luminosity of the entire protocluster, $L_{IR}$∼ 8 × 10$^{13}$L$_\odot$ (Strandet et al. 2016; Miller et al. 2018; Hill et al. 2020). The factor 1.7× [Cɪɪ] line emission from the Core already indicates that there exists a significant amount of gas (possibly between these systems) around the Core members. Additional faint objects revealed in the deep ALMA measurements by Hill et al. (2020) resulted in detecting multiple new objects and fainter emission surrounding some of the objects within the Core (<1 Jy km s$^{-1}$) that were not originally detectable by Miller et al. (2018). The fainter objects only increased the total ALMA-detected [Cɪɪ] line intensity by about only ∼10–15 Jy km s$^{-1}$ for the same area covered by APEX. This suggests that the 1.7× [Cɪɪ] emission in the Core does not come from the recently detected faint [Cɪɪ]-emitting systems, but from an ICM/CGM phase, with the excess [Cɪɪ] emission even much more pronounced in the North component.

### 3.2. First [Oɪɪɪ] 88 μm line detection in a z > 1 protocluster

The [Oɪɪɪ] 88 μm line was first detected in the early Universe in two composite AGN/starburst systems (Ferkinhoff et al. 2010). In the local Universe, it is thought to be powered solely by O star radiation fields (e.g., Cormier et al. 2012; Lambert-Huyghe et al. 2022 and references therein), even as its critical density ($n_{e,crit}$ ∼ 510 cm$^{-3}$) and low excitation potential $E_{10}/k_B$ ∼ 163 K allows it to remain luminous for lower temperature and densities of ionised gas beyond the immediate vicinity and/or inside young/compact Hɪɪ regions (within those this line is insensitive to their much higher temperatures and will be nearly LTE-excited for $n_e \gtrsim 10^3$ cm$^{-3}$). The [Oɪɪɪ] emission in the Core has comparable line fluxes and profile as the [Cɪɪ] line (Figure 1), as observed in other known cases for dusty star-forming galaxies (De Breuck et al. 2019). The [Oɪɪɪ] line has a similar luminosity to [Cɪɪ] despite a factor of three smaller area covered in the pointing[6] (Fig. 1,A.1), with $L_{[OIII]}/L_{[CII]}$= 1.3 and $L_{[OIII]}$= (8.3 ± 1.7) × 10$^{10}$ L$_\odot$; yet, as noted in Sect. 1 these two lines trace different gas phases altogether. We can nevertheless use the [Oɪɪɪ] frequency-integrated flux density $F_{10}$ to obtain the minimum amount of ionised hydrogen gas from:

$$M_{min}(H^+) = F_{10} \left[ \frac{4\pi D_L^2}{(g_1/Z(T_e) e^{-E_1/k_B T_e}) A_{10} h\nu_{10}} \right] \frac{m_H}{X(O^{++})} \quad (1)$$

where $n_1 = N_1/N_{tot} = [g_1/Z(T_e)] e^{-E_1/k_B T_e}$ is the occupation number of the upper level (for LTE), $Z(T_e)$ the 3-level partition function, $A_{10}$ = 2.7 × 10$^{-5}$ s$^{-1}$ (Einstein coefficient for spontaneous emission), $\nu_{10}$=3393.00624 GHz the line frequency, and $X(O^{++})$ the fractional abundance of oxygen in the doubly ionised state. For hot plasma: $e^{-E_1/(k_B T_e)} = e^{-163/T_e(K)} \sim 1$, and for all Oxygen atoms in the doubly-ionised state (i.e. $X(O^{++}) \sim X(O) = 5.9 \times 10^{-4}$; assumed Milky Way value) we obtain

$$M_{min}(H^+) \sim 2.6 \times 10^{-4} X(O)^{-1} \left( \frac{L_{10}}{L_\odot} \right) M_\odot \quad (2)$$

($L_{10}$ the line luminosity), which yields: $M_{min}(H^+)$∼3.7×10$^{10}$ M$_\odot$. Sub-thermal line excitation, cooler plasma, optical depth effects reducing the observed [Oɪɪɪ] line intensity, or Oxygen atoms at a lower ionisation state, will all increase the actual M(H$^+$). This minimum ionised gas mass already amounts to 30% of the reported molecular gas masses reported for the sum of all objects within the APEX beam (Hill et al. 2020, see Table 3). The question then arises whether the high [Oɪɪɪ] line (and the large implied ionised gas mass) can be explained as the outcome of the total IR-derived SFR for the Core. APEX [Oɪɪɪ] observations cover less than a third of the 23 Core objects. Estimates of the IR-derived SFR from these objects result in SFR ∼2500 M$_\odot$ yr$^{-1}$, while the larger Core area within 300 kpc (proper distance), with 23 objects, has a total IR-derived SFR of ∼ 5000 M$_\odot$ yr$^{-1}$(Hill et al. 2020).

## 4. Discussion

Here we discuss the significance of these single-dish [Cɪɪ] and [Oɪɪɪ] line detections. We first derive an independent SFR estimate from the [Oɪɪɪ] line luminosity. As mentioned before, the [Oɪɪɪ] line is typically expected to trace star-forming regions in the vicinity of intense ionisation sources due to the necessity of 35.1 eV photons required to create O$^{2+}$. In deriving such a SFR and comparing to the known IR-derived SFR we can explore whether or not the [Oɪɪɪ]-deduced SFR is high enough to consider the possibility of significant non-stellar contributions to the observed [Oɪɪɪ] line luminosity from other sources in or around individual star-forming galaxies within SPT2349-56.

It can be shown that for only stellar types O5.5V or above, the amount of ionising flux is enough to keep He doubly ionised (requires 24.6 eV) throughout the Strömgren sphere (e.g., Draine 2011), so the presence of such stars is essential for attributing the observed [OIII] luminosity to star formation activity. First we estimate the number of representative O5.5 class V stars, whose typical bolometric luminosity is $10^{5.41}$ $L_\odot$ (see Table 1 in Martins et al. 2005). Metal lines are the primary coolants of photoionised gas, emitting about 1% of the incident bolometric luminosity. Therefore, we use $\epsilon_O = 1\%$ to upconvert the observed [OIII] luminosity to the expected bolometric luminosity of the stars responsible for keeping the gas ionised, $L_{[OIII]}/\epsilon_O = N_{O5.5V} \times L_{O5.5V}$. This results in $\sim$32.3 million O5.5V stars. Using STARBURST99 (Leitherer et al. 1999) stellar population synthesis models, we find that a continuous star formation episode with SFR=1 $M_\odot$ yr$^{-1}$ sustained over 6–8 Myrs results in about $10^{4.35}$ O stars. It should be noted that these models count all stars $\gtrsim$15.6 $M_\odot$ as O-stars, so we calculate the integral over the IMF used for SB99[7] with limits from $M_{low}$=34.4$M_\odot$ to $M_{up}$=100$M_\odot$ and find the fraction of stars O5.5V or greater, $f_{(\%>O5.5V M_\odot)} = 28.57\%$. We therefore derive a scaling between a SFR=1 $M_\odot$ yr$^{-1}$ and $N_{O5.5V}=f_{(\%>O5.5V M_\odot)}N$(O stars, 1$M_\odot$ yr$^{-1}$)$\sim$6400, as shown in this independent [OIII] 88 $\mu$m-derived equation for the SFR using these known properties of O-stars and the STARBURST99 models closest to solar metallicity ($Z = 0.02$), thus

$$\frac{\text{SFR}_{\text{OIII}}}{M_\odot \text{ yr}^{-1}} = \frac{L_{\text{OIII}}}{\epsilon_O L_{O5.5V}} \times \frac{1}{f_{(\%>O5.5V M_\odot)} N_{\text{(O stars, }1M_\odot \text{ yr}^{-1})}}, \quad (3)$$

which gives SFR$_{\text{OIII}}\sim$5050$M_\odot$ yr$^{-1}$, exceeding the 2500 $M_\odot$ yr$^{-1}$ value estimated for this area of the protocluster (Hill et al. 2020). The OIII-deduced SFR will be even higher if the assumed metallicity is lower, further exaggerating the discrepancy.

---

[5] These objects are not within the pointing of the primary beam (∼2–3") of APEX.

[6] Due to the differences in the beam size between 358 GHz and 640 GHz.

The previous equation has demonstrated that the OIII-deduced SFR may be equal to or even higher than the total IR-derived SFR. Therefore, we further provide an independent estimate of the formation rate of only O stars from the [OIII]-deduced $H^+$ mass reservoir, requiring two assumptions: a) this reservoir consists solely of young ultra-compact HII (UCHII) regions around O5-O7 stars (those whose high ionisation parameter U can maintain the $O^{++}$ ionisation state), b) these stars are expected to be replenished in a steady-state manner by an equal amount of molecular gas mass (copious amounts of which have been detected in this protocluster; (e.g. Zhou et al. 2025)). Using the turbulence-regulated SF theory (Krumholz & McKee 2005), this replenishment occurs by collapsing dense gas cores on timescales: $\Delta t_* = f_* \Delta t_{ff} = (f_*/4)\sqrt{3\pi/(2G\langle\rho\rangle)}$, thus:

$$\frac{\text{SFR}_{\text{O stars}}}{M_\odot \text{ yr}^{-1}} \sim \frac{\epsilon_* M_{\min}(H^+)}{\Delta t_*} \sim 818 \left(\frac{\epsilon_*}{f_*}\right)\left(\frac{\langle n \rangle}{\text{cm}^{-3}}\right)^{1/2} \quad (4)$$

within the protocluster area sampled by the APEX beam. From turbulence-regulated SF theory: $f_*=(\text{SFR}_{ff})^{-1}\sim$70-180, where SFR$_{ff}\sim$0.0056-0.014 is the normalized SFR per free-fall time $\Delta t_{ff}$ (for core virial parameters of $a_{vir}$=1.3-5, see Equations 21 and 30 in Krumholz & McKee (2005)), while $\epsilon_*$=$M_O/M_{UCHII}$ is the O star mass per ionised gas mass of its UCHII region. For molecular cloud core densities $\langle n \rangle = (1-5)\times 10^4$cm$^{-3}$ (typical of the dense SF molecular gas), and $\epsilon_*\sim$0.65 (for a UCHII size of r$\sim$0.5 pc, and using the $n_e$-(2r) relation from Kim & Koo (2001) (Figure 9), and a mean mass of $M_O$=40 $M_\odot$ for O6-O7 spectral types), we obtain an SFR$_{\text{O stars}}\sim$(300–1700) $M_\odot$ yr$^{-1}$. This already amounts to $\sim$12%-68% of the total, IR-derived SFR computed for this area of the protocluster (assuming a standard stellar IMF), while for a lower X(O$^{++}$)=[O$^{++}$/H$^+$]=0.3$\times$(O)=1.77$\times 10^{-4}$ abundance inside the HII regions (appropriate for O5.5 stars with H$^+$ and He$^+$ throughout their volume)[8] M(H$^+$)$_{\min}\sim$1.2$\times 10^{11}$ $M_\odot$, raising the implied SFR$_{\text{O stars}}$ to $\sim$(1000-6000)$M_\odot$ yr$^{-1}$, which is at least $\sim$65% and can even surpasses the IR-derived total SFR (an obvious impossibility). The strength of two independent arguments that estimate the OIII-deduced SFR suggest the possibility for non-stellar contributions to the observed [OIII] line luminosity or top-heavy stellar IMFs (a possible outcome of HEP-dominated ISM; Papadopoulos et al. 2011), important issues to be addressed by future studies and spatially resolved analyses.

The APEX [OIII] observations covered the region of the Core, encompassing only three of the brightest galaxies initially discovered by Miller et al. (2018) (objects 'C1,C3,C7'; Hill et al. 2020), Object 'C1' (band 10 undetected) and 'C3' (band 10 detected) are among the brightest reported individual [CII] emitters (Miller et al. 2018; Hill et al. 2020), yet 'C3' is associated with a known radio-loud AGN system composed of 'C3,C6,C13' (all of which are detected in band 10 and indicate a strong dust heating source). Subsequent X-ray observations have indicated that 'C1' and 'C6' host AGN with roughly Compton-thick gas ($N_H\sim 10^{24}$ cm$^{-2}$; Chapman et al. 2024; Vito et al. 2024). The contribution of these galaxies to the APEX-observed [OIII] luminosity is of course uncertain, as is whether any extended [OIII]-emitting ICM phase of the protocluster exists. Moreover the systems detected in the ALMA band 10 measurements ('C3,C6,C13'; Fig. A.1) are dusty starbursts, and could have significant optical depths even in the far-IR. Indeed extreme dusty/merger starbursts like Arp 220 have been found optically thick out to 450$\mu$m (Papadopoulos et al. 2010) and even 2.6 mm (Scoville et al. 2017), while *The Red Radio Ring*, a strongly lensed DSFG at z=2.6, seems to be optically thick out to 200 $\mu$m (Harrington et al. 2019). Needless to say that any dust optical depth corrections or low-density ionised gas phase (sub-thermally exciting the [OIII] line) in the protocluster can only raise the corresponding M(H$^+$), making its attribution solely to HII regions of O stars even more problematic (as these are produced by a standard IMF).

$L_{\text{OIII}}/L_{\text{CII}}$ ratios of >1 have been observed before for high redshift starburst galaxies at z$\sim$6-9 (Harikane et al. 2020), higher than local dwarf galaxies and local ULIRGs (Fujimoto et al. 2024 and references therein). In such high redshift systems where O-stars seem to be in high abundance, we would expect to have a richer release of O as opposed to C (which comes from the lower mass and thus longer lived stars). This affects the C/O abundance ratio making the so-called $\alpha$-enhanced ISM. In an $\alpha$-enhanced ISM one can also expect to have high $L_{\text{OIII}}/L_{\text{CII}}$ (Bisbas et al. 2024, 2025), however it is not clear whether the ISM/ICM power sources are dominated by the SF-produced far-UV photons or are HEP-dominated instead. Recently the environment around a known dusty star-forming galaxy at $z = 2.8$ was imaged by JWST, revealing 2.5 kpc filamentary structures of [OIII]$\lambda\lambda$4959,5007 line emission extending out to 60 kpc beyond the host system (Peng et al. 2025), exactly tracing the previously imaged Lyman-$\alpha$ emission, indicating shock-heated CGM. Solimano et al. (2025) reported that radiative shocks were insufficient to power the [OIII]$\lambda\lambda$4959,5007 line emission in a $z \sim 4.5$ protocluster, and explored the possibility of outflows and AGN in powering these CGM-scale [OIII]. Harikane et al. (2025) have recently analysed both singly and doubly ionized optical and far-IR oxygen lines, suggesting the far-IR emission lines may arise predominantly from relatively low-density gas (electron density $n_e \sim 100 \text{cm}^{-3}$). Moreover, (far-UV)-photon deficient thermal ISM states have been found dominating over large molecular gas reservoirs even inside extreme dusty starbursts like NGC 6240 (Papadopoulos et al. 2014), a distinct possibility also for the ICM/CGM in galaxy clusters (Ferland et al. 2008, 2009; Lim et al. 2017). The initial conditions of star formation are strongly impacted by the dominant ISM heating source, and so will be the resulting in-situ (galaxies) and ex-situ (CGM/ICM) stellar IMF that builds up the stellar mass in the Universe.

---

[7] https://massivestars.stsci.edu/starburst99/figs/fig38.html

[8] The value 0.3 is representative of the distribution between 0.2-0.6 in Figure 14 of Amayo et al. (2021).

The significance of the [Cɪɪ] excess (1.4× in the Core and 4× in the North compared to the star-forming galaxies) is that it likely traces a diffuse CGM or a pre-heated proto-ICM, which may play a critical role in sustaining the high SFR observed. The larger [Cɪɪ] excess in the North region could reflect variations in environmental conditions, but we need further investigation to address the origin (see e.g. Fujimoto et al. 2019). Without spatially resolved observations of these emission lines, we can only speculate but not definitely confirm the source of ionising photons at this point. It is assumed that extended [Oɪɪɪ] may be less expected than extended [Cɪɪ] given the higher IP. While previous works have explored the possibility of extended [Cɪɪ] halos around individual galaxies and merging systems using ALMA (e.g. Fujimoto et al. 2019; Ginolfi et al. 2020), more recent single-dish observations have begun to uncover even more spatially extended [Cɪɪ] in protocluster systems (e.g. De Breuck et al. 2022). We have evidence that [Oɪɪɪ], having a much higher IP, has a similar distribution, as inferred from the line width of [Cɪɪ]. This begs for an explanation for what causes the widespread [Oɪɪɪ] line emission. Is widespread star formation enough to explain this, or is it required to invoke explanations like shock-driven excitation and/or widespread AGN activity (considering there may be higher AGN activity in protoclusters (Vito et al. 2024; Shah et al. 2024), including SPT2349-56)? This frames the high-stakes in using reliable ionised and neutral gas phase line ratio diagnostics for determining the thermal and ionisation states in extraordinary systems like the SPT2349-56 protocluster in the Early Universe.

## 5. Conclusions and outlook

Two new APEX single-dish measurements of the [Cɪɪ] line in the SPT2349-56 protocluster reveal $\sim 1.7\times$ the amount of [Cɪɪ] line emission than previously measured with ALMA 12-m observations of the Core, and the [Cɪɪ] detection from the North reveals $\sim 4\times$ additional [Cɪɪ] line luminosity than detected with deep ALMA 12-m observations – both suggesting the presence of additional amounts of ICM/CGM gas. An APEX [Oɪɪɪ] line measurement towards the Core revealed the presence of a large reservoir of $O^{++}$, corresponding to a massive $H^+$ gas reservoir. The latter is difficult to reconcile solely as the product of star formation in this intensely star forming protocluster, unless non-stellar sources (HEPs and/or strong shocks) contributing to the CGM/ICM ionisation, or non-standard (top-heavy) stellar IMFs are considered. It is thus an imperative that the multi-phase CGM/ICM reservoir in this and other such extraordinary systems is studied both in its ionised and molecular/neutral-atomic gas phase, using appropriate line ratios – reliably measured from scales of individual galaxies up to the large scales of their CGM/ICM distributions. This is impossible with interferometers alone, given the short u-v spacing filtering-out the large scale emission, yielding biased views of both the true mass of the CGM/ICM components and their thermal and chemical states. Our work demonstrates the necessity of overcoming these short-spacing issues in interferometric data, by also obtaining measurements of such far-IR fine-structure lines with single-dish facilities. This further motivates the development of future facilities like the Atacama Large Aperture Submillimeter Telescope (i.e. AtLAST Mroczkowski et al. 2025), reaching up to the necessary high frequency ranges, without filtering out line emission over large scales.

*Acknowledgements.* The authors would like to pay our respects to the passing of German Astronomer, Dr. Karl Menten: https://www.mpifr-bonn.mpg.de/announcements/2025/1. He was the founder of APEX, had an enormous influence in ensuring the development of the Atacama Large Millimeter/submillimeter Array, and has left a scientific legacy for years and years to come.
Ανδρος επιφανους πας ουρανος σκεπη εστι.
The authors would like to thank the referee for their thoughtful comments and constructive review. The authors also thank Rob Ivison, Laya Ghodsi, Nick Foo, as well as Dazhi Zhou, Scott Chapman and Nikolaus Sulzenauer, for helpful comments and discussions. We would also like to thank all of the APEX team of operators, observers and staff for making these successful observations, including Juan-Pablo Perez-Beapuis, Manuel Merello, Felipe MacAuliffe, Claudio Agurto, Mauricio Martinez, Francisco Azagra, Pablo Garcia and Macarena Lopez. PPP would like to thank the ESO visitorship program that hosted him in the period November-March 2024-25, as well as Thorsten Naab and Bo Peng at MPA for the hospitality and the intense discussions. This work was supported by the European Southern Observatory through available travel funds for Fellows, which has been greatly appreciated by KCH. AWSM acknowledges the support of the Natural Sciences and Engineering Research Council of Canada (NSERC) through grant reference number RGPIN-2021-03046 and the ESO Visitor Program. The data were collected under the Atacama Pathfinder EXperiment (APEX) Project, led by the Max Planck Institute for Radio Astronomy at the ESO La Silla Paranal Observatory. This paper makes use of the following ALMA data: ADS/JAO.ALMA2024.0.01465.S. ALMA is a partnership of ESO (representing its member states), NSF (USA) and NINS (Japan), together with NRC (Canada), NSTC and ASIAA (Taiwan), and KASI (Republic of Korea), in cooperation with the Republic of Chile. The Joint ALMA Observatory is operated by ESO, AUI/NRAO and NAOJ. Reproduced with permission from Astronomy & Astrophysics, © ESO

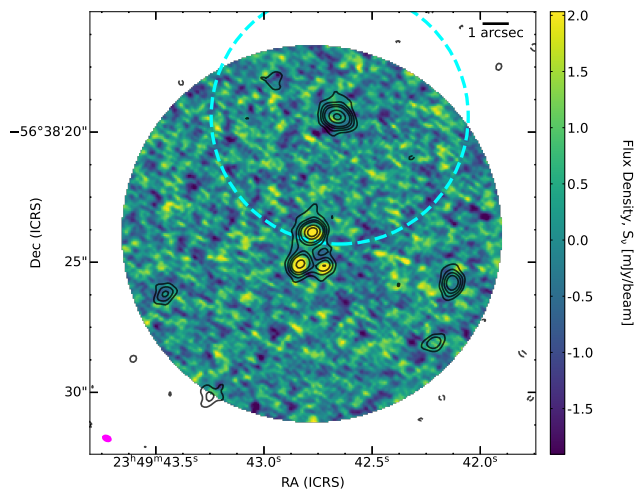

**Fig. A.1.** The ALMA band 10 continuum image of the Core region imaged beyond the primary beam of ~7" to explore emission from bright sources (B10 synthesized beam (magenta) shown in the bottom left corner). The APEX measurement of the [OIII] emission line comes from the cyan circle. Black contours show the band 6 dust continuum detections at signal-to-noise ratio levels of -3,3,6,10,15.

**Table A.1.** ALMA band 10 continuum detections

| Object name | RA J2000 | DEC J2000 | Flux density mJy |
|---|---|---|---|
| C3 | 23:49:42.774 | -56:38:23.875 | 8.5 ± 0.8 |
| C6 | 23:49:42.837 | -56:38:25.070 | 7.0 ± 0.8 |
| C13 | 23:49:42.719 | -56:38:25.246 | 3.7 ± 0.6 |

**Notes.** The aperture based photometry used a circular diameter centered on the known positions ('Object Name' from Hill et al. 2020). A 0.7″ diameter was used for both C3 and C6, and a 0.54″ diameter aperture was used for C13. All other objects have $3\sigma$ upper limits of ~2 mJy beam$^{-1}$.

## Appendix A: ALMA band 6 and 10 continuum

The band 10 ALMA receiver is a dual-sideband receiver that requires 90 degrees Walsh switching to split the sidebands and corresponding spectral windows in order to maximize the continuum bandwidth to ~ 8 GHz (Maud et al. 2022, 2023). This dataset was observed using the band-to-band calibration method of observing a differential gain calibrator at a lower frequency than the high frequency tuning in order to scale the higher signal-to-noise solutions to higher frequencies before eventually applying to the target source (Asaki et al. 2020b,a, 2023; Maud et al. 2020, 2022, 2023). The array configuration for band 10 reaches a field of view of ~7", maximum recoverable scale of ~3.2" and synthesized beam $\theta$ ~0.19" with Briggs weighting and a robust parameter of 0.5. For the archival ALMA band 6 data, the synthesised beam (with Briggs robust weighting factor = 0.5) is $\theta$ ~ 0.45". The field of view is 25", with a maximum recoverable scale of ~6.5".

## Appendix B: Novel APEX observations compared with LABOCA sub-mm maps

APEX spectral line pointed observations used wobbler switching, which only chops in Azumith, with a 1.5 Hz rate and offset of 50". Each scan consisted of a hot/sky/cold calibration, followed by up to 10 subscans of 20 s per on-source integration time. Wobbler-switching spectral line focus observations were performed every 2-4 hr and pointing checks every 1-3 hr, de-pending on frequency, using planets (Mars, Saturn) or standard APEX calibrators (e.g. 'pi1-Gru', 'R-Dor', 'IRAS15194-5115'). The central position for the wobbler-switching is set to an amplitude of 50", which results in an off position that is 100" from the target, as shown in the Fig B.1. A representative set of 'off' positions is shown for the APEX/SEPIA observations with respect to APEX LABOCA submm continuum maps (Hill et al. 2020).

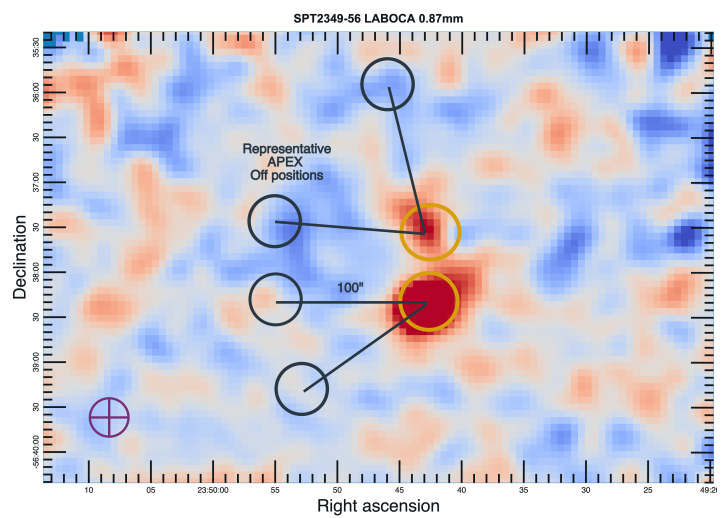

**Fig. B.1.** Representative 'off' positions (black circles) are shown for the APEX/SEPIA observations of the Core and North (orange circles), overlaid on LABOCA submm continuum emission (magenta beam shown in bottom left).

Our result independently verifies the APEX total power measurements first reported by Strandet et al. (2016), and recently highlighted by Zhou et al. (2025) in the context of excess CO(4-3) emission. We note that the pointing in Strandet et al. (2016) was identical as ours, and still agrees with our novel measurements of the Core using SEPIA345 with better receiver baseline stability. This consistency between measurements remains despite using a completely different instrument (i.e. the decommissioned FLASH instrument (Klein et al. 2014), with the observations in Strandet et al. (2016) having taken place before the complete refurbishment of the primary mirror for APEX in 2017.

If we consider the global dust emitting region, previous LABOCA measurements had measured more 870 μm flux density than reported with ALMA (Miller et al. 2018; Hill et al. 2020). The elongated structure that extends to the North is where we find a relative excess of 4× amount of [CII] emitting gas compared to the one galaxy, yet extended dust from LABOCA is likely diffuse, warm and optically thin. We therefore consider dust attenuation of the [OIII] line from any large-scale dust emission to be negligible. Still, this suggests there could be additional line emission that are resolved out by interferometers like ALMA.